# Equity in College Physics Student Learning: a Critical Quantitative Intersectionality Investigation

Ben Van Dusen & Jayson Nissen
California State University Chico

We investigated the intersectional nature of race/racism and gender/sexism in broad scale inequities in physics student learning using a critical quantitative intersectionality. To provide transparency and create a nuanced picture of learning, we problematized the measurement of equity by using two competing operationalizations of equity: *Equity of Individuality* and *Equality of Learning*. These two models led to conflicting conclusions. The analyses used hierarchical linear models to examine student's conceptual learning as measured by gains in scores on research-based assessments administered as pretests and posttests. The data came from the Learning About STEM Student Outcomes' (LASSO) national database and included data from 13,857 students in 187 first-semester college physics courses. Findings showed differences in student gains across gender and race. Large gender differences existed for White and Hispanic students but not for Asian, Black, and Pacific Islander students. The models predicted larger gains for students in collaborative learning than in lecture-based courses. The Equity of Individuality operationalization indicated that collaborative instruction improved equity because all groups learned more with collaborative learning. The Equality of Learning operationalization indicated that collaborative instruction did not improve equity because differences between groups were unaffected. We discuss the implications of these mixed findings and identify areas for future research using critical quantitative perspectives in education research.

## Introduction

The science and education communities have highlighted the need to better support the learning of diverse student populations (NRC, 2010; 2013; NSF, 2017; 2019; NASEM, 2019). In their review of discipline-based education research (DBER), the National Research Council (NRC, 2012, pg. 136-137) stated that while, "[DBER] clearly indicates that student-centered instructional strategies can positively influence students' learning… Most of the studies the committee reviewed were not designed to examine differences in terms of gender, ethnicity, socioeconomic status, or other student characteristics." We addressed this gap in the literature by using a critical quantitative intersectionality framework (Jang, 2018; Covarrubias, 2011; Covarrubias & Vélez, 2013) to problematize measuring equity and explore the intersectional nature of race/racism and gender/sexism in introductory college physics courses.

University physics courses have transformed over the prior decade to emphasize collaborative learning activities over lecture-based instruction (NRC, 2012). These changes have moved the focal point of classrooms from the instructor to the students and in doing so reshaped the classroom power structures. Consequently, collaboration-based curriculums improve student learning compared to lecture-based teacher-centered models of instruction (Hake, 1998; NRC, 2012; Freeman et al., 2014). Research on the impact of curricular transformations, however, has largely ignored their impacts on students from marginalized groups in physics (NRC, 2012; McCullough, 2018; NASEM, 2019). The need to assess the impact of these curriculums across races and genders is highlighted by the fact that these curriculums were developed at research-intensive institutions with and for a highly selective set of students (Kanim & Cid, 2017).

Existing research on student learning in physics that disaggregates its findings across demographic groups has focused primarily on gender differences (McCollough, 2018). Most quantitative studies that examined the role of race/racism in physics student learning included data from only one institution that was research intensive (Watkins, 2010; Kost, Pollock & Finklestein, 2009) or majority minority (Brewe et al., 2010). These quantitative studies all combined minority groups, found that post-course differences were accounted for by pre-course differences, and did not examine the intersectionality of race and gender.

We are unaware of any publications that quantitatively examined intersectionality across race/racism and gender/sexism in physics. Intersectionality is a theoretical and methodological framework for investigating why, how, and in what situations students from multiple intersecting marginalized groups have different experiences. Intersectionality can inform student experiences and outcomes across multiple axes of power and privilege in STEM fields (Ireland et al., 2018; Nash, 2008; Crenshaw, 1991). For example, Black women face marginalization from White men, White women, and Black men in their physics education (Dortch, 2017). The studies examining the intersection of physics student races and genders have used qualitative methods to investigate the experiences of women of color (Fries-Britt, Younger, & Hall, 2010; Fries-Britt, Johnson, & Burt, 2013; Hyatar-Adams, 2018; Rosa & Mensah, 2016; McGee & Bentley, 2017; Holmes, 2013; Ong, 2005; Rahm & Moore, 2016). The present study differs from these prior studies by using a critical quantitative intersectionality framework to operationalize equity and explore the intersectional experiences of college physics students using statistical models.

The lack of quantitative research on race/racism, gender/sexism, and their intersectionality in physics partially driven by the lack of diversity in physics. While this study examines the broader population of students who take introductory physics courses rather than just those majoring in physics, the difficulty of collecting data on students from marginalized groups is reflected in their shares of physics degrees earned. Females earned 20% of the physics bachelor degrees in 2016 (AIP Statistical Research Center, 2019a), Hispanics and African-Americans earn a combined 10% of the physics bachelor degrees from 2014 to 2016 (AIP Statistical Research Center, 2019b), and African-American, Hispanic, and Native-American females earned a combined 1.7% of physics degrees from 2003 to 2013 (Merner & Tyler, 2017).

With such low rates of representation, it is difficult to collect sufficiently large datasets to analyze outcomes for students doubly-underrepresented by gender and race. To overcome this difficulty, we analyzed data from a national database of science course outcomes on the Learning About STEM Student Outcomes (LASSO) platform (LA Alliance, 2019). The LASSO national database included data from 13,857 students in 187 first-semester college physics courses. 2,196 students in the dataset identified as being from a group marginalized by gender and race.

To contextualize the associations between lecture and collaborative learning in physics courses with the intersection of race/racism and gender/sexism, we examined learning through the lens of equity. Equity is a term that has been used to mean many things, even within the education research community (Espinoza, 2007). Critical quantitative intersectionality calls for questioning and testing the measures and analytical practice to offer competing practices that better describe the experiences of students from marginalized groups (Stage, 2007). Rodriguez and colleagues (2012) problematized using a singular implicit definition of equity when examining physics student learning. We use two competing operationalizations of equity to interpret our findings and create a richer picture of the experiences of students from marginalized groups. The first operationalization of equity (Equality of Learning) compares the physics learning of students from marginalized groups to those of White male students. The second operationalization of equity (Equity of Individuality) drops the use of White male students as the normative standard and just examines the learning of marginalized students across course types. Our Theoretical Framework sections further describes these operationalizations.

## Research Questions

This study investigates the intersectionality of race/racism and gender/sexism in physics courses that primarily use either lecture or collaborative learning and their associations with equity in student learning. This study includes five research questions. The first two questions investigated intersectionality in student physics preparation and learning.

1. To what extent does the intersection of race/racism and gender/sexism predict physics student preparation?
2. To what extent does the intersection of race/racism and gender/sexism predict physics student learning?

The next two research questions investigated Equality of Learning.

3. What differences in student learning emerge across the intersection of race/racism and gender/sexism in physics courses that are lecture-based, if any?
4. To what extent are differences in student learning across the intersection of race/racism and gender/sexism smaller in physics courses that use collaborative learning?

The fifth research question investigated Equity of Individuality when transforming from lecture-based to collaborative physics courses.

5. To what extent do collaborative learning physics courses increase learning across the intersection of race/racism and gender/sexism compared to those that are lecture-based?

## Literature Review

While research on inequities in STEM education may come from a desire to improve outcomes for students from marginalized groups, the work often lends itself to interpreting differences as deficiencies in students from marginalized groups (Gutierrez, 2008; Gutierrez & Dixon-Romain, 2011; Hazari & Potvin, 2005). For example, someone could interpret a finding that female physics students have smaller average gains than their male peers in multiple ways. Interpreted through a deficit perspective, these findings could support arguments like the one then Harvard president Larry Summers's made claiming that women did not perform as well in math and science because of biological differences (Goldenburg, 2005). Interpreted through a critical perspective, these findings support calls for transforming the institutional, departmental, and classroom power structures to better serve students from marginalized groups. These same issues arise with what we often see as neutral terminology, such as underrepresented minority. While it is factually accurate that Black, Hispanic, and female students are underrepresented in physics, the term leaves it to the reader to interpret whether their underrepresentation is caused by deficits in the students or the systems they are embedded in.

### Gender/sexism in Physics

Many investigations of equity in STEM do not distinguish between the fields within STEM (Cheryan et al., 2017). For example, in Talking About Leaving, Seymour (2000) investigated why men and women leave STEM majors, and they tended to treat STEM as a single domain and did not focus on differences across the STEM discipline. Cheryan et al. (2017) argued that representation and experiences of women varies across STEM disciplines and needs to be looked at within each discipline. They reviewed the literature on gender differences across the STEM domains and found that masculine cultures, gender differences in self-efficacy, and a lack of early educational experiences in the disciplines explained the lower rates of participation for women in physics, computer science, and engineering compared to biology, chemistry, and mathematics. Cheryan and colleagues (2017) results show that inequities vary across disciplines.

Significant research efforts have focused on understanding the underrepresentation of women in physics, which has remained at approximately 20% for the last 40 years (AIP Statistical Research Center, 2019a). This work is highlighted by a 2016 special collection in Physical Review Physics Education Research titled *Gender in Physics* (Brewe & Sawtelle, 2016) and Madsen, McKagan, and Sayre's (2013) review of 26 studies on gender differences for conceptual learning in introductory physics courses. In first semester physics courses, Madsen et al. (2013) found that in the 26 studies they reviewed the male students' average pretest scores were always higher than female students' average pretest scores (13% weighted average) and in most cases male students' average posttest scores were also higher (12% weighted average). Most of the 26 studies did not find statistically significant differences in the learning across genders. Madsen and colleagues concluded that the studies did not identify a single factor or solution for the gender differences, but that the gender differences are likely due to a combination of factors.

Some studies have identified course transformations associated with decreases in gender gaps on research-based assessments. Lawrenz et al. (2009) investigated the relationships between teaching with a reform curriculum and gender differences in conceptual learning. They operationalized equity as being when there is equivalent learning across groups of students while controlling for their pretest scores. We refer to this operationalization of equity as *Equality of Learning* and address it further in the Operationalizing Equity section. They found that boys (0.44 s.d.) and White students (0.88 s.d.) had higher posttest scores than girls and students from minority groups. Use of the reformed curriculum was associated with higher posttest scores (0.33 s.d.) and moderated the gender difference (0.22 s.d. reduction) but did not moderate race/ethnicity differences. Their findings indicated the student characteristics they controlled for had a much stronger relationship with student achievement than teacher moderated variables.

Lorenzo et al. (2006) examined the gender gap in three physics classroom environments they describe as traditional, interactive engagement, and full interactive engagement. They found that there were consistent gender gaps in pretest scores ranging from 8.5-15% but as the level of interactive engagement increased the gender gaps in student posttest scores decreased. In the interactive engagement and full interactive engagement courses the raw gains for female students were 2.8%-7.4% larger than those of the male students. Lorenzo and colleagues concluded that interactive engagement reduces the gender gap in introductory physics courses. The investigation, however, was critiqued by Rodriguez et al., 2012 for its quantitative analysis and implicit definition of equity, which we discuss below.

**Race/racism in Physics**

Few studies have investigated race/racism in college physics (McCullough, 2018). The quantitative studies that have tend to combine students into two groups: majority (White and Asian) and underrepresented minority (Black, Hispanic, and all others). Some studies only look at differences after instruction while other studies control for preexisting differences. Watkins (2010), Brewe et al. (2010), and Kost, Pollock and Finkelstein (2009) all used research-based assessments as a pretest and a posttest for conceptual knowledge. All three studies found the differences in conceptual knowledge after instruction between majority and marginalized students were explained by preexisting differences. In contrast, Van Dusen et al. (2016) and Van Dusen and Nissen (2018) found that differences in conceptual knowledge increased from pretest to posttest and the differences on the posttest were not explained by preexisting differences. One exception to combining minority groups is Hazari, Sadler, and Sonnert's (2013) investigation of science identity. They found that college students who are Hispanic women/Latina reported much weaker science identities than any other group.

While these quantitative studies tended to combine marginalized groups, some qualitative studies have focused on the lived experiences of students of color (Fries-Britt, Younger, & Hall, 2010), Black students (Fries-Britt, Johnson, & Burt, 2013), Black women (Hyatar-Adams, 2018; Rosa & Mensah, 2016; McGee & Bentley, 2017; Holmes, 2013), and women of color (Ong, 2005) in physics. Students of color experience their race as a salient component of their physics education. They are often ignored and avoided by their fellow peers and faculty members, they are dissuaded from pursuing STEM degrees by faculty members, and excluded from insider know-how needed to succeed in their education (Seymour, 2000; McPherson, 2017; Johnson 2001, 2005; Dortch, 2017; Ong, 2005; McGee and Bentley, 2017). These negative experiences occurred less frequently for male Black students at HBCU's (Dortch, 2017). However, Black women faced exclusion at HBCU's and primarily White institutions because they face marginalization from White men, White women, and Black men (Dortch, 2017).

**Operationalizing Equity in Physics**

Quantitative studies in college physics courses seldom explicitly operationalize equity. Most quantitative PER studies that examine equity implicitly operationalize it to mean different demographic groups have either the same average gains/effect sizes (Lorenzo et al., 2006; Brewe et al., 2013; Van Dusen et al., 2016) or posttest scores (Day et al., 2016; Watkins, 2010; Van Dusen and Nissen, 2018). Researchers have referred to these types of equity as "equity for equal potential" (Espinoza, 2007) when controlling for background preparation or "equity of fairness" (Lee, 1999; Rodriguez et al., 2012) when ignoring background preparation. Some science education researchers, however, have problematized the literature's use of terms such as equity. Lee (1999) points out that while research often uses equity and equality interchangeably, there are important distinctions between them. Lee describes equity as, "associated with fairness and justice, whereas equality is associated with sameness or an absence of difference" (1999, p. 89).

Rodriguez et al. (2012) demonstrated the need to operationalize equity when supporting claims about students from groups marginalized in physics by reexamining the claims of Lorenzo et al. (2006) about gender equity in physics classrooms. Using Lorenzo and colleagues reported results that fully interactive classes achieved gender equity, Rodriguez et al. (2012) applied a more robust quantitative analysis technique (Cohen's *d* instead of normalized learning gain) (Nissen et al., 2018) and examined the results through three operationalizations of equity: 1) equity of parity - equal posttest scores across groups regardless of pretest scores, 2) equity of fairness - equal gains across groups regardless of pretest scores, and 3) equity of individuality - an intervention improves outcomes for a specific marginalized group. In their original analysis, Lorenzo et al. concluded that, "in the fully interactive courses, the gender gap is entirely eliminated" (2006, p. 121). After reanalyzing the data, Rodriguez and colleagues found nuanced variations in which classroom settings met the different operationalizations of equity. They concluded that, contrary to Lorenzo and colleagues' claim, the fully interactive classroom did not entirely achieve gender equity. Rodriguez et al. also concluded that while the predominant trend in equity research is to compare gaps in group performance on posttests, it is important that researchers consider other operationalizations of equity, such as equity of individuality, that may lead to designing different interventions and learning environments.

## Theoretical Framework
**Critical Quantitative Intersectionality**

Critical perspectives, such as critical race theory (CRT) (Ladson-Billings & Tate, 1995), are grounded in the assumption that race is a salient feature of the American education system that demands explicit attention. Studies framed in CRT highlight the diversity of students' identities and the complexity of their interactions with the racist structures present in our educational systems (Sleeter & Delgado, 2003; Ladson-Billings, 2003). Intersectionality (Crenshaw, 1991; Collins and Bilge, 2016) argues that understanding the complex experiences of student and educators requires accounting for the multiple axes of social division (e.g., race, gender, class, job title) that interact and influence one another. Critical Quantitative Intersectionality (CQI) provides a framework for applying critical theory and intersectionality, which have been used primarily for qualitative studies (Jang, 2018; Stage, 2007), to quantitative studies. CQI reframes the question from 'what's more important: race, gender, or class?' to 'how can we examine unique constellations of race, gender, class social location as categories of experience in a given educational context (Lopez et al., 2018)?' CQI positions us to, "*use data to represent educational processes and outcomes on a large scale to reveal inequities and to identify social or institutional perpetuation of systematic inequities in such processes and outcomes*"; and to "*question the models, measures, and analytic practices of quantitative research in order to offer competing models, measures, and analytic practices that better describe experiences of those who have not been adequately represented*" (Stage, 2007, p.10-11).

Investigating the learning of students from marginalized groups presents inherent difficulties. Marginalized groups in physics are typically underrepresented minorities, which makes it challenging to collect sufficient data for quantitative analysis. These challenges are exacerbated for studies of intersectionality which examine within-group differences for subsets of students from marginalized groups, such as distinguishing between Black men and Black women. Quantitative criticalists overcome this challenge by collecting large-scale datasets with enough data to model the relationships between students' intersectional identities and their learning outcomes. The recent emergence of large-scale databases of university science student data (e.g., LASSO, 2019; DataExplorer, 2019; E-CLASS, 2019) have made it easier to obtain the statistical power needed to model the impacts of intersecting race/racist and gender/sexist power structures.

Most quantitative publications on STEM student equity try to take a neutral stance and let the numbers speak for themselves (Covarrubias & Vélez, 2013). We, however, take a critical perspective that explicitly problematizes power structures and their roles in perpetuating inequities. To avoid our findings being interpreted from a deficit perspective, we often used advocative terms (e.g., gender/sexism, race/racism, and marginalized) over the neutral terms commonly used in the literature (e.g., gender, race, and underrepresented minority).

**Operationalizing Equity**

In our analysis, we follow the advice of Rodriguez et al. (2012) and Stage (2007) by offering two competing operationalizations and associated measures of equity to better describe the experiences of students from marginalized groups. Our two operationalizations are grounded in the literature but we have renamed them to ease the reader's interpretation and to align with Lee's (1999) definition of equity and equality. The two operationalizations of equity we use in our analysis are: 1) *Equality of Learning* and *2) Equity of Individuality*.

*Equality of Learning* is achieved when students from different gender and racial groups learn equivalent amounts. This perspective has been called "equity for equal potential" (Espinoza, 2007) and is related to "equity of fairness" (Lee, 1999; Rodriguez et al., 2012). Equality of Learning ensures that students who start the semester similarly prepared attain the same level of achievement. This is a commonly used type of equity, but it can be problematic in several ways. First, achieving Equality of Learning maintains gaps in initial preparation caused by student's opportunity gaps or educational debt (Ladson Billings, 2006). Second, it perpetuates positioning White males as the ideal students who other groups should strive to emulate (Dixon-Roman, 2011; Gutierrez, 2008) and perpetuates a majoritarian interpretation (Covarrubias et al., 2018).

*Equity of Individuality* is achieved when an intervention improves the outcomes of students from marginalized groups (Rodriguez et al., 2012). This perspective gets away from making comparisons with White, middle-class students and what Gutierrez & Dixon-Roman (2011) refers to as "gap gazing" to focus instead on research and interventions designed to advance the needs of marginalized groups. Gutierrez (2008) argues that the focus on achievement gaps supports a deficit model of students from marginalized groups. While Equity of Individuality no longer relies on the majoritarian comparison to majority groups (Covarrubias et al., 2018), it has its own shortcomings. By only focusing on one population of students, achieving Equity of Individuality can still exacerbate gaps between student populations.

By using two different operationalizations of equity to interpret our findings, we seek to create a more complete picture of the experiences of students from marginalized groups. We use both Equality of Learning and Equity of Individuality to problematize using a single measure of equity by comparing and contrasting the results for each. We hope it will help foster a larger conversation within the DBER community about what equality and equity are, how they are measured, and how they are achieved.

## Methods

Our multi-institution, multi-level dataset came from the Learning About STEM Student Outcomes (LASSO) platform (LASSO, 2019) database. Using R, we cleaned the dataset, created ten complete datasets using multiple imputation, developed hierarchical linear models (HLM) for the ten datasets, and analyzed and pooled the results from the HLM models. Supplemental Figure S1 shows our workflow for the data collection and analysis.

### Data Collection and Processing

The data for this study comes from two research-based assessments commonly used in introductory college physics courses (Madsen et al., 2016), the Force Concept Inventory (FCI; Hestenes & Swackhamer, 1992) and the Force and Motion Conceptual Evaluation (FMCE; Thornton & Sokoloff, 1998). Both the FCI and FMCE assess core concepts in first semester physics courses and focus on forces and motion. Both assessments have had significant validation work (Dietz et al., 2012; Traxler et al., 2018; Thornton et al., 2009) and researchers have typically found absolute gains from pretest to posttest ranging between 10% to 30% (Hake, 1998; Caballero et al., 2012; Rodriguez, 2012).

We accessed student and course data through the Learning About STEM Student Outcomes (LASSO) platform (Van Dusen, 2018). The LASSO platform is an online platform that collects large-scale, multi-institution data by administering, scoring, and analyzing pretest and posttest research-based assessments. The LASSO platform makes an anonymized version of its database of course and student data (for those who consent to share it) available to researchers.

The data in this study came from 15,267 students in 201 courses from 32 institutions. For each student, the data included their gender, race, ethnicity, pretest score, posttest score, time spent taking the assessment, and which course they were in. For each course, the platform provided us the assessment used and whether it used collaborative learning.

To clean the data, we applied student level filters and then course level filters. We removed student level data if the gender data or response to whether they were retaking the course was missing. We removed the pretest or posttest score if the student took less than 5 minutes on the assessment or completed less than 80% of the questions. Five minutes provided a reasonable minimum amount of time for a student to complete the CI while reading and answering each question. If a student had neither a pretest nor a posttest score after these filters, we removed their data. We removed courses with less than 40% student participation on either the pretest or posttest from the data and courses with fewer than either 8 pretests or 8 posttests. Courses with very low participation and low numbers of students likely represented unreliable data. The size of the dataset after each step in the filtering process is shown in Table 1. After filtering, 58% of the students had matched pretest and posttest scores, which fell in the range of typical participation rates in the literature (Nissen, Donatello, & Van Dusen, in press).

We calculated pretest and posttest scores using total percentage correct of all the items.

Table 1.
*Size of the dataset after each step of filtering.*

|  | Initial | Missing Data | Time and Completion | Course Filters |
|---|---|---|---|---|
| Institutions | 32 | 32 | 32 | 31 |
| Courses | 201 | 201 | 201 | 187 |
| Students | 15,267 | 15,193 | 14,676 | 13,857 |

**Handling missing data with hierarchical multiple imputation**

After cleaning the data, we used hierarchical multiple imputation (HMI) with the hmi (Speidel et al., 2018) and mice (van Buuren & Groothuis-Oudshoorn, 2010) packages in R-Studio V. 1.1.456 to address missing data. HMI is a principled method for maximizing statistical power by addressing missing data while accounting for the hierarchical structure of the data (Allison, 2002; Buhi et al., 2008; Manly & Wells, 2015; Schafer, 1999). HMI addresses missing data by (1) imputing each missing data point $m$ times to create $m$ complete datasets, (2) independently analyzing each dataset, and (3) combining the $m$ results using standardized methods (Drechsler, 2015). Multiple imputation does not produce specific values for missing data; it uses all the available data to produce valid statistical inferences (Manly & Wells, 2015).

Our HMI model included variables for the concept inventory used, pretest and posttest scores, gender, and the type of instruction in the course. The data collection platform (LASSO) provided complete data sets for the concept inventory variables, student demographics, and instruction type. The 42.1% rate of missing data (12.6% on pretests plus 29.5% on posttests) in this study was within the normal range for PER studies using pretests and posttests (Nissen, Donatello, & Van Dusen, in press). The HMI produced 10 imputed complete datasets. We analyzed all 10 imputed data sets and combined the results by averaging the test statistics (e.g., model coefficients) and using Rubin's Rules to combine the standard errors for these test statistics (Schafer, 1999). All HMI assumptions were satisfactorily met for all the HMI analyses. For more information on HMI, Schafer (1999) and Manly & Wells (2015) provide overviews and Nissen et al. (in press) examine its use specifically in DBER.

**Descriptive Statistics and Contexts**

We used student's self-reported demographic data collected through the LASSO platform to categorize their genders and races/ethnicities. As our data had few students select transgender or "other" for their gender (N=105), we included them with the larger group of female students

4marginalized in physics. Race/ethnicity categories included Black, Hispanic, Asian, Hawaiian or Pacific Islander (Pac. Islander), other, or White based on student responses to the demographics questionnaire. The other category included students who selected other, Alaskan Native or Indian American, or did not respond. We included the Alaskan Native and Indian American students in the other category because they represented a small sample of students (N=61). We identified students with multiple racial identities by the identity with the smallest sample size to make the race/ethnicity categories independent of one another to simplify the model and preserve statistical power. For example, a student who identified as both Black and Hispanic was included as a Black student because that was the smaller sample size. With the exception of Hispanic students, the number of students with multiple races/ethnicities was small. Of the 2,162 students included as Hispanic in our model, 985 identified as both White and Hispanic.

The data set included 187 courses: 153 courses used collaborative instruction, had 11,740 students, and mean gains of 20.9% ($\sigma$ = 19.5%); 34 courses used lecture instruction, had 2,117 students and mean gains of 15.7% ($\sigma$ = 18.3%). The FMCE was used in 48 of the courses. The FCI was used in 139 courses. We calculated descriptive statistics for the dataset to characterize differences between the mean pretest scores, posttest scores, and gains across student demographics (Table S1). Because the sample included fewer students in lecture courses (Table 2), several of the samples for different race/ethnicities are very small: in particular there are only 64 Black and 15 Pacific Islander or Hawaiian students in courses that used lecture.

The final sample included students and courses from 31 institutions. The highest degree granted by these institutions varied and included 20 institutions that granted doctorates, 6 that granted master's degrees, 3 that granted bachelor's degrees, and 2 that granted associate degrees.

Table 2
*Sample sizes across instruction type, gender, and race/ethnicity.*

| Race | Collaborative | | Lecture | |
|---|---|---|---|---|
| | Male | Female | Male | Female |
| Asian | 811 | 635 | 96 | 75 |
| Black | 169 | 180 | 28 | 36 |
| Hispanic | 1,192 | 568 | 204 | 198 |
| Pac. Islander | 69 | 42 | 5 | 10 |
| Other | 497 | 357 | 92 | 95 |
| White | 4,791 | 2,429 | 593 | 685 |

**Model Development**

To investigate student learning, we developed two sets of models to predict student pretest scores and gains (*posttest - pretest*). As the mean scores for the pretest (34.6%) and posttest (55.7%) in our dataset indicate that the floor and ceiling effects were limited, no transformation of the data was required prior to analysis (Day et al., 2016). Pretest and gains are represented as a percentage correct. Our models were all 2-level hierarchical linear models with student data in the first level and course data in the second level. Using hierarchical linear models allowed us to account for the nested nature of our dataset (Van Dusen & Nissen, in press). We explored whether including a 3rd level for institution improved our models, but we found that it increased the total variance and did not substantively change the model coefficients. We developed the models and pooled the results for the imputed datasets using the mitml (Grund et al., 2016) and lme4 (Bates et al., 2014) packages in R.

We developed our models for student pretest and gain scores through a series of additions of variables. The first model we developed was the unconditional model, which predicts the student gains without level-1 or level-2 predictor variables. The unconditional model allowed us to calculate the intraclass correlation coefficient (ICC) by comparing the course and student level variance in the HLM model. The ICC showed course-level effects accounted for 21.5% of the variation in student pretest scores and 15.9% of the variation in student gains and fell above the heuristic threshold of 5%, indicating that HLM was necessary.

We evaluated each model on whether including additional variables improved the model's goodness of fit. Goodness of fit can be assessed through the examination of several statistics such as Akaike information criterion (AIC), Bayesian information criterion (BIC), and variance explained. As there is currently no agreed upon way to pool the AIC or BIC statistics for multi-level models across multiply imputed datasets, we used variance explained to select our final model. To identify the simplest model that accounted for the most variance, we only included variables that improved the student- and/or course-level variance explained by at least 1%.

**Pretest Models.** Our pretest model development included variables for the instrument used ($FMCE_j$), gender ($Female_{ij}$), race ($Asian_{ij}$, $Black_{ij}$, $Hispanic_{ij}$, $PacIslander_{ij}$, and $OtherRace_{ij}$), course type ($Lecture_j$), interaction effects, and whether a student had previously taken the course ($Retake_{ij}$). The variables for whether a student had previously taken the course ($Retake_{ij}$) increased the variance of the model and was removed. A table for the development of the pretest models is included in the supplemental material (Table S2). The final pretest model used the following equations (we combined 11 level 2 equations that are isomorphic for brevity):

**Level 1 Pretest Equation**

$StudentPre_{ij} = \beta_{0j} + \beta_{1j} * Female_{ij} + \beta_{2j} * Asian_{ij} + \beta_{3j} * Black_{ij} + \beta_{4j} * Hispanic_{ij} + \beta_{5j} * PacIslander_{ij} + \beta_{6j} * OtherRace_{ij} + Female_{ij} * (\beta_{7j} * Asian_{ij} + \beta_{8j} * Black_{ij} + \beta_{9j} * Hispanic_{ij} + \beta_{10j} * PacIslander_{ij} + \beta_{11j} * OtherRace_{ij}) + r_{ij}$

**Level 2 Pretest Equations**

$\beta_{0j} = \gamma_{00} + \gamma_{01} * FMCE_j + \mu_{0j}$
$\beta_{1j} = \gamma_{10} + \gamma_{11} * FMCE_j$
$\beta_{(2-11)j} = \gamma_{(2-11)0}$

**Gain Models.** Our gain model development included all the variables used in the pretest models plus a pair of variables for pretest scores ($StudentPreCen_{ij}$ and $CoursePreCen_j$). $StudentPreCen_{ij}$ was a student-level variable for a student's pretest score and $CoursePreCen_j$ was a course-level variable for a course's average pretest score. Both of the pretest score variables were centered for the gain model (details below). The inclusion of the $StudentPreCen_{ij}$ variable makes the gain models isomorphic with models that use posttest score with the exception of a shift in the intercept value. As the use of gain and posttest score lead to the same findings, we opt to use gain as we feel the model is more easily interpreted. $CoursePreCen_j$ and the instrument used ($FMCE_j$) variables were found to not improve the fit of the models and were removed. The failure of the instrument variable ($FMCE_j$) to explain variance in the gain model when it did the pretest model indicates that the differences in scores across instruments are similar on the pretest and posttest and make no differences in the gains. $Retake_{ij}$ and $StudentPre_{ij}$ variables improved the models fit, and we included them in the final models. We include a table for the development of the gain models in the supplemental material (Table S3). We identified 2 gain models to use in our analysis. The first gain model does not include a variable for course type, allowing us to assess the aggregated gains across all course types. The second gain model disaggregates predicted gains across lecture-based and interactive physics courses. The final gain models used the following equations (we combined 13 level 2 equations that are isomorphic for brevity):

**Level 1 Gain Equation Independent of Course Type**

$Gain_{ij} = \beta_{0j} + \beta_{1j} * StudentPreCen_{ij} + \beta_{2j} * Retake_{ij} + \beta_{3j} * Female_{ij} + \beta_{4j} * Asian_{ij} + \beta_{5j} * Black_{ij} + \beta_{6j} * Hispanic_{ij} + \beta_{7j} * PacIslander_{ij} + \beta_{8j} * OtherRace_{ij} + Female_{ij} * (\beta_{9j} * Asian_{ij} + \beta_{10j} * Black_{ij} + \beta_{11j} * Hispanic_{ij} + \beta_{12j} * PacIslander_{ij} + \beta_{13j} * OtherRace_{ij}) + r_{ij}$

**Level 2 Gain Equations Independent of Course Type**

$\beta_{0j} = \gamma_{00} + \mu_{0j}$

$\beta_{(1-13)j} = \gamma_{(1-13)0}$

**Level 1 Gain Equation Accounting for Course Type**

$Gain_{ij} = \beta_{0j} + \beta_{1j} * StudentPreCen_{ij} + \beta_{2j} * Retake_{ij} + \beta_{3j} * Female_{ij} + \beta_{4j} * Asian_{ij} + \beta_{5j} * Black_{ij} + \beta_{6j} * Hispanic_{ij} + \beta_{7j} * PacIslander_{ij} + \beta_{8j} * OtherRace_{ij} + Female_{ij} * (\beta_{9j} * Asian_{ij} + \beta_{10j} * Black_{ij} + \beta_{11j} * Hispanic_{ij} + \beta_{12j} * PacIslander_{ij} + \beta_{13j} * OtherRace_{ij}) + r_{ij}$

**Level 2 Gain Equations Accounting for Course Type**

$\beta_{0j} = \gamma_{00} + \gamma_{01} * Lecture_j + \mu_{0j}$

$\beta_{(1-13)j} = \gamma_{(1-13)0}$

For ease of interpretation, we group mean centered student pretest scores ($StudentPreCen_{ij}$) in the gain model. This has two effects on the models. First, the intercept ($\beta_{0j}$) represents the predicted gain for a student who had the average pretest score in their course (i.e., group). Second, the coefficient for $StudentPre_{ij}$ informs the relationship between an above (or below) average pretest in a course and the predicted gain. We used group mean centering because it is generally recommended in the literature (Bauer & Curran, 2005), particularly in our case where an uncentered model predicts values for a pretest of zero, which is unlikely and more difficult to interpret. Group mean centering $StudentPreCen_{ij}$ kept the intercept consistent across the models. As part of a sensitivity analysis, we also ran our final models with student prescores grand mean centered, but we did not find any meaningful differences in the coefficients of interest. We left all other variables in the model uncentered.

All assumptions were met for all the final models. Additional information about the assumption checking are included in the Assumption Checking section and Figure S1 of the Supplemental Materials.

## Findings

In the findings section we analyzed the output of our pretest and gain models. We included the coefficients created by each model in Table 3. We used these models to examine the role of intersectionality in predicting student preparation (measured by pretest scores) and learning (measured by student gain). We then analyzed the predicted gains for evidence of student equity.

Table 3
*Coefficients for the pretest, gains independent of course type, and gains accounting for course type models. The models for pretest and gains independent of course type inform the status and intersectionality of race/racism and gender/sexism in physics student's conceptual learning and preparation. The model of gains accounting for course type differentiates between lecture-based and collaborative instruction to inform the investigation of equity across course types. Variable subscripts are consistent with model of gains accounting for course type (with the exception of the FMCE variables).*

| Fixed Effect | Pretest | | | Gains Independent of Course Type | | | Gains Accounting for Course Type | | |
|---|---|---|---|---|---|---|---|---|---|
| | Coeff. | S.E. | p-value | Coef. | S.E. | p-value | Coef. | S.E. | p-value |
| Intercept, $\gamma_{00}$ | 44.11 | 0.7 | <0.001 | 19.83 | 0.83 | <0.001 | 20.42 | 0.87 | <0.001 |
| Lecture, $\gamma_{01}$ | - | - | - | - | - | - | -3.06 | 1.61 | 0.057 |
| FMCE, $\gamma_{01}$ | -10.45 | 1.31 | <0.001 | - | - | - | - | - | - |
| Student Prescore, $\gamma_{10}$ | - | - | - | -0.33 | 0.01 | <0.001 | -0.33 | 0.01 | <0.001 |
| Retake, $\gamma_{20}$ | - | - | - | -2.77 | 0.6 | <0.001 | -2.74 | 0.64 | <0.001 |
| Female, $\gamma_{30}$ | -11.68 | 0.48 | <0.001 | -2.17 | 0.5 | <0.001 | -2.16 | 0.5 | <0.001 |
| Female * FMCE, $\gamma_{11}$ | 2.62 | 0.69 | <0.001 | - | - | - | - | - | - |
| Asian, $\gamma_{40}$ | -3.14 | 0.66 | <0.001 | -2.42 | 0.68 | <0.001 | -2.42 | 0.68 | <0.001 |
| Black, $\gamma_{50}$ | -11.52 | 1.35 | <0.001 | -7.43 | 1.62 | <0.001 | -7.42 | 1.62 | <0.001 |
| Hispanic, $\gamma_{60}$ | -8 | 0.58 | <0.001 | -3.91 | 0.73 | <0.001 | -3.91 | 0.73 | <0.001 |
| Pac Islander, $\gamma_{70}$ | -5.89 | 2.12 | 0.006 | -5.41 | 2.45 | 0.029 | -5.43 | 2.45 | 0.029 |
| Other Race, $\gamma_{80}$ | -5.96 | 0.85 | <0.001 | -4 | 0.93 | <0.001 | -3.99 | 0.93 | <0.001 |
| Female * Asian, $\gamma_{90}$ | 3.42 | 0.098 | 0.001 | 1.2 | 1 | 0.238 | 1.18 | 1 | 0.238 |
| Female * Black, $\gamma_{100}$ | 6.33 | 1.87 | 0.001 | 3.08 | 2.07 | 0.144 | 3.05 | 2.07 | 0.144 |
| Female * Hispanic, $\gamma_{110}$ | 5.17 | 0.92 | <0.001 | -0.84 | 1.15 | 0.468 | -0.84 | 1.15 | 0.468 |
| Female * PacIslander, $\gamma_{120}$ | 4.69 | 3.29 | 0.154 | 2.46 | 3.64 | 0.5 | 2.46 | 3.64 | 0.5 |
| Female * OtherRace, $\gamma_{130}$ | 4.02 | 1.28 | 0.002 | 2.17 | 1.39 | 0.125 | 2.16 | 1.39 | 0.125 |

**Intersectionality in Student Preparation.** The pretest model predicted meaningful and statistically significant differences in pretest scores across genders, races, and instruments (Figure 1 and Table 3). Large racial differences existed with the highest pretest scores predicted for White male students and the lowest predicted pretest scores for Black female students. Gender differences existed within each racial group with higher pretest scores predicted for male students than female students. Interaction effects between gender and race varied across the racial groups. The model predicted the largest gender difference for White students on the FCI (11.7 percentage points). The interaction coefficients for race and gender were positive for women of color and so moderated the predicted gender differences to range between -5.35 and -8.26 percentage points for students of color.

The model predicted that the pretest scores on the FMCE would be 10.45 percentage points lower than on the FCI. Adding interaction effects between race and FMCE did not improve the variance explained. The interaction effect for gender and FMCE, however, indicated that gender gaps were 2.62 percentage points smaller on the FMCE than the FCI.

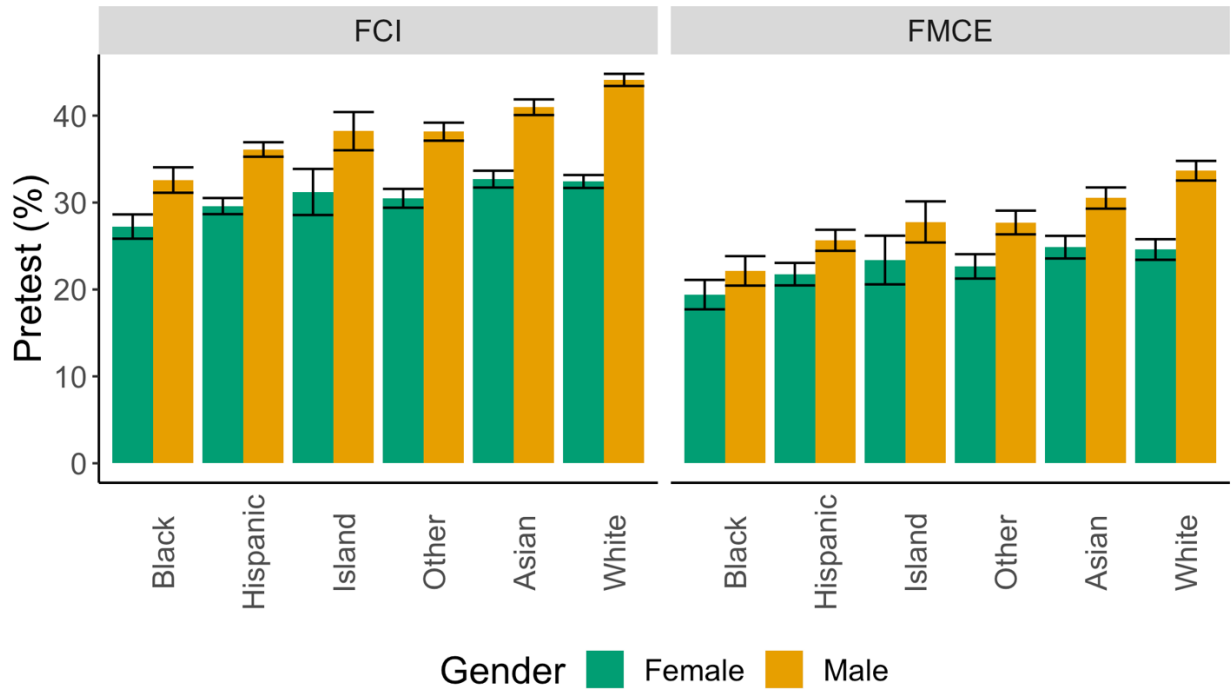

**Figure 1.** Predicted pretest scores disaggregated by race, gender, and instrument. Error bars represent +/-1 S.E.

**Intersectionality in Student Learning.** The aggregated gains model predicted meaningful and statistically significant differences in gains across genders and races (Figure 2). Further, the model found that the relationship between gender and predicted gains varied across racial groups. Specifically, after accounting for differences in pretest scores White and Hispanic female students were predicted to have statistically significantly lower gains than their male peers. Asian, Black, or Pacific Island female students, however, were predicted to have gains that are more similar to their male peers. Black and Pacific Island female students were even predicted to slightly outperform their male peers. None of the differences in predicted gains across genders for students who are Asian, Black, or Pacific Island were statistically significantly different. This lack of statistically significant differences come from smaller predicted differences and sometimes larger standard errors because of smaller sample sizes for some marginalized groups.

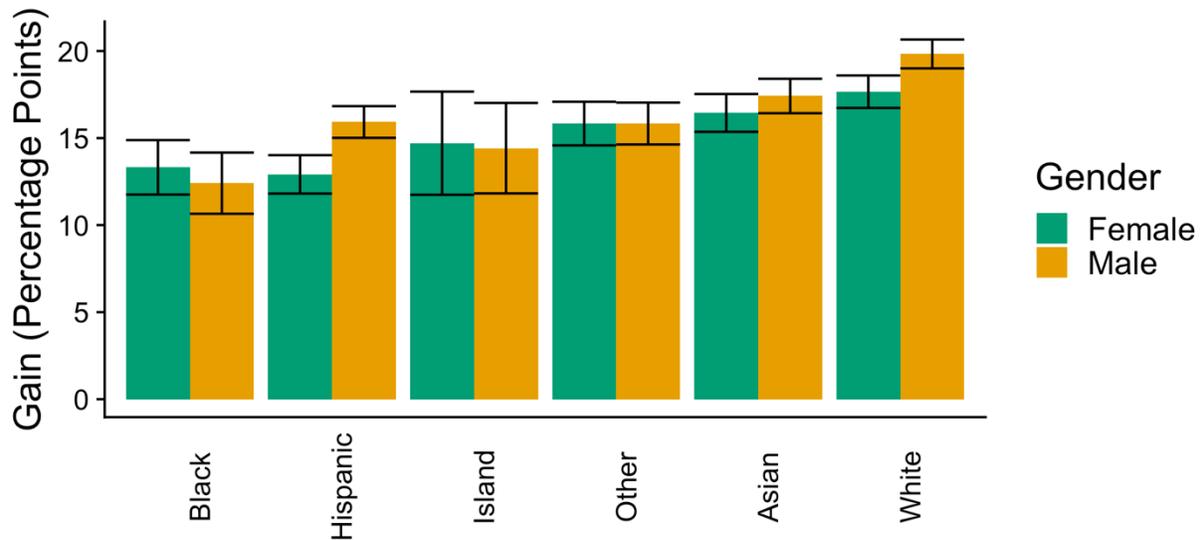

**Figure 2.** Predicted group gains independent of course type (Model 5) disaggregated by demographics groups while controlling for pretest scores. Error bars represent +/-1 S.E.

**Equality of Learning.** To examine whether lecture-based or collaborative courses achieved Equality of Learning, we calculate the differences in gains for each demographic group relative to White males in the same course type. Since the disaggregated gain model showed that including an interaction effect for course type with gender and race variables did not improve the fit of the model, we concluded that the differences in gains between student groups was the same in lecture-based and collaborative courses. Figure 3 shows the differences in group gains in either course types compared to white males in the same course type, as predicted by the disaggregated gain model. The model predicts that the gains for students from each marginalized group are smaller than those of their White male peers even though the model treats them as beginning a course with equal pretest scores. The predicted differences in gains between White males and every other group are statistically significantly different, except for Pacific Island females who had the lowest representation in the data and therefore the largest confidence intervals. The predicted gap in gains in both course contexts is smallest for White females (-2.17 percentage points) and largest for Black males (-7.42 percentage points). To contextualize the differences in predicted gains, White female students gained 88% and Black male students gained 57% as much as their equally well prepared White male peers. These gaps are both statistically and meaningful significant and indicate that Equality of Learning was not achieved in either course context.

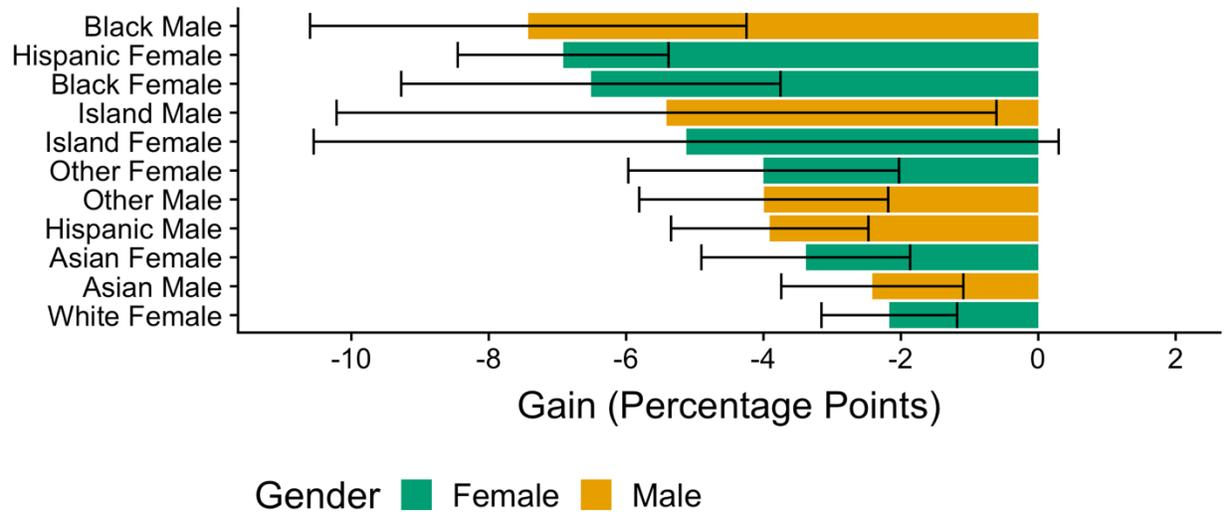

**Figure 3.** Differences in predicted gains independent of course type (Model 5) between each demographic group and White males (gain$_{group}$ - gain$_{white\ male}$) controlling for pretest scores. Error bars represent 95% confidence intervals.

**Equity of Individuality.** To examine whether courses that used collaborative learning achieved Equity of Individuality, we examined the impact of course type on predicted gains for each demographic group in the gain model accounting for course type. The model predicts that all demographic groups will have gains that are 3.06 percentage points larger in collaborative courses than lecture-based courses. To contextualize the magnitude of these improvements in predicted gains, we calculated the ratio of predicted gains for each marginalized group in collaborative courses with the same group in traditional courses (gain$_{collab.}$ / gain$_{trad.}$). While the raw predicted improvement in gains in collaborative learning courses over lecture-based courses is the same for each group, because each group has a different predicted gain in lecture-based courses the proportional magnitude of the change varies. Improvements in predicted gains for marginalized groups in collaborative courses range from 20.1% for White females to 30.8% for Black males (Table 4). These improvements in gains are meaningful and indicate that Equity of Individuality was achieved in collaborative courses.

Table 4.
*Percent increase in learning gains in collaborative courses compared to lecture-based courses by race and gender.*

| Race | Female | Male |
|---|---|---|
| White | 20.1% | 17.6% |
| Asian | 21.9% | 20.5% |
| Hispanic | 29.3% | 22.7% |
| Other | 22.9% | 22.9% |
| Island | 25.0% | 25.6% |
| Black | 28.3% | 30.8% |

**Discussion**

This study's findings that non-White and female physics students have smaller gains than their White male peers controlling for pretest scores is in alignment with prior studies. This study, however, makes several novel contributions to the literature. First, this is one of the first studies to examine equity in physics student learning across a wide range of college and course types. The breadth of the contexts allowed us to develop comparative models while strengthening the generalizability of our findings. Second, our large sample size provided sufficient statistical power to disaggregate students across both gender and racial groups. This allowed us to examine the relationships between gender/sexism, race/racism, and their intersectionality with student learning in introductory college physics courses. Third, we explicitly operationalized equity in two different ways to provide a broader perspective on the relationship between using collaborative learning and equity while critically examining the implications of these two measures of equity. In what follows, we discuss how our findings answer our five research questions.

R.Q. 1: To what extent does the intersection of race/racism and gender/sexism predict physics student preparation?

Our pretest model indicates that there are meaningful differences in prior preparation between gender and race groups. The predicted pretest score for White men are meaningfully and statistically significantly higher than for every other racial and gender group. Within each racial group, the men had higher predicted pretest scores than females. The largest such gender gap in predicted pretest scores was for White students on the FCI. Gender gaps were 2.62 percentage points smaller on the FMCE. The smaller gender gaps in pretest scores on the FMCE versus the FCI is aligned with prior researchers' findings of gender bias on the FCI (Traxler et al., 2018). Besides having higher predicted pretest scores, men had a much larger range of predicted pretest scores across racial groups (11.52 percentage points) than women (5.46 percentage points). Our models cannot identify the cause of men's higher and more variable predicted pretest scores; however, they speak to the cumulative effects of systemic sexism and racism in suppressing the preparation of marginalized physics students.

R.Q. 2: To what extent does the intersection of race/racism and gender/sexism predict physics student learning?

Our gain model independent of course type, indicates that the average gains are larger for White males students than any other demographic group, even after controlling for difference in preparation. Examining the intersectionality of gender/sexism and race/racism, we see that the biggest differences in predicted gains is between racial groups. The model only identified gender difference for White and Hispanic students. We note that half of the Hispanic students identified as White and our model failed to account for that specific intersectional identity. As Jang (2018) found differences in academic outcomes across ethnic groups frequently combined under Asian, our results point to the possibility of differences within the Hispanic ethnicity. For example, it could be that the gender differences for Hispanic students are being disproportionately driven by data from White Hispanic students. The nuanced differences in predicted gains across racial and gender groups shows the importance and complexity of modeling the impact of overlapping systems of sexism and racism.

R.Q. 3: What differences in student learning emerge across the intersection of race/racism and gender/sexism in physics courses that are lecture-based, if any?

To answer our research questions about equality and equity, we examined our gain model accounting for course type, which predicted student gains for either lecture-based or collaborative learning courses after accounting for pretest scores. The model showed that there were persistent gaps in gains between White males and every other demographic group in lecture-based courses. The gaps in gains compared to White males ranged from -2.2% for White females to -7.4% for Black males. To put that in context, even after controlling for differences in pretest scores the predicted gain for White males in lecture-based courses (17.4%) is 75% larger than those of Black males (10.0%). This provides empirical evidence that our lecture-based physics courses are creating meaningful gaps in student gains and do not lead to Equality of Learning.

R.Q. 4: To what extent are differences in student learning across the intersection of race/racism and gender/sexism smaller in physics courses that use collaborative learning?

In developing our gain model accounting for course type we found that creating a model that allowed for the interaction of demographic variables with course type did not improve the fit of the model. This means that our data indicated that the gaps in student gains across demographic groups were not related to the type of instruction used in the course. The model showed students in courses that used collaborative learning had larger predicted gains than their peers in lecture-based courses. However, the absolute differences in the gains between demographic groups were not improved by collaborative instruction. As with lecture-based courses, we found no evidence that collaborative instruction achieved Equality of Learning. Instead, we found that collaborative instruction and lecture-based instruction increased the disparities in physics knowledge between White male students and all other students.

R.Q. 5: To what extent do collaborative learning physics courses increase learning across the intersection of race/racism and gender/sexism compared to those that are lecture-based?

When we stop using White males as the normative group and examine the models for Equity of Individuality in physics courses that use collaborative learning, we come to a different conclusion. From this perspective, the predicted improvement in gains for students from marginalized groups in collaborative courses over lecture-based courses represents an improvement in equity. That the use of collaborative learning also improved outcomes for white males does not diminish the value of the improvement in gains for students from marginalized groups. The absolute increase of 3.1 percentage points in gains represents a proportionally larger and impactful improvement for the groups with smaller gains in lecture-based courses. For students from marginalized groups, the predicted gains in lecture-based courses ranged from 9.9 to 15.2 percentage points. The 3.1 percentage point increase in collaborative courses represents a 20.1% to 30.8% increase in gains for students from marginalized groups. These results indicate that collaborative learning supports Equity of Individuality across marginalized student groups.

**Conclusions and Implications**

Students from marginalized groups enter physics courses less prepared on average than their White male peers. This result is consistent with a lack of early educational experiences in part explaining the underrepresentation of women in physics, engineering and computer science compared to mathematics, biology, and chemistry (Cheryan et al., 2017). These differences in physics preparation are not because of any deficiencies in the students themselves, but a lifetime of science and math experiences that have preferentially catered to White male students.

Introductory college physics courses perpetuate these inequities in preparation by creating environments in which White male students have the largest gains. These differences in gains act to accentuate the differences in physics knowledge between White men and students from marginalized groups. This result shows that addressing inequities requires creating more changes to the power structures in college physics education than only integrating collaborative learning activities. This is not surprising when considering classroom power structures. Collaborative learning redistributes some of the power from the instructor to the students. The culture of physics, however, is often hypermasculine and cutthroat (Cheryan et al., 2017; Seymour & Hewitt, 1994). If left unchecked, the distribution of power amongst students benefits students who identify with that masculine culture and students whose privileged background has better prepared them to compete for grades in physics courses. While students may learn more on average when the instructor releases some of their power, these environments will not lead to more equitable outcomes unless they also address existing power differentials between students.

The associations between equity in student learning and collaborative learning were mixed; collaborative courses achieved Equity of Individuality but not Equality of Learning. Both operationalizations of equity can provide insight into equity, but they also create a tension worth examining. Equity scholars (Gutierrez & Dixon-Roman, 2011) proposed the use of Equity of Individuality to get away from positioning White male students as the ideal state for comparisons. Focusing on gaps between majority and marginalized groups has an assimilationist goal that frames marginalized groups as being deficient. However, equity occurs at multiple levels such as within a course, across a set of courses, and at the national level. Multiple measures and models of equity are necessary to examine it from multiple perspectives.

Comparisons are a powerful tool for understanding the impacts of classroom interventions and transformations on student learning. As each operationalization of equity provides useful perspectives for understanding student outcomes, a single operationalization cannot provide a broad enough description of equity to fully inform policy and practice. The stark differences in conclusions about equity that researchers can draw from analyzing the same dataset highlights the importance of explicitly operationalizing of equity (Rodriguez et al., 2012). We also call on researchers to include enough detail in their results to enable others to evaluate their findings with different operationalizations of equity.

Interest convergence, a central tenet of Critical Race Theory, posits that interventions to support equity will only be implemented when they benefit the group in power. Collaborative instruction supports students from all demographic groups in learning more. However, if collaborative instruction methods are disproportionately used at affluent, research-intensive, primarily White institutions, and in courses that disproportionately serve male students then they perpetuate the systemic racism and sexism in physics. The data in this study show that students in lecture-based courses disproportionately come from marginalized groups compared to students in collaborative courses (66% versus 50%). This inequity likely occurs at a national level in the United States. Two-year colleges enroll a disproportionate number of students from marginalized groups and hold a huge potential for increasing the number and diversity of STEM students (Bahr et al., 2017; NASEM, 2019). Two-year colleges tend to spend less per student and are less likely to have the resources to implement collaborative learning. We do not know the extent to which collaborative instruction and other research-based instructional strategies are used in physics courses at two-year colleges because almost no research looks at physics instruction in two-year colleges (Kanim and Cid, in review; NRC, 2012). Large-scale studies looking at both the equity of interventions on student outcomes and the equity in how they are distributed across courses and institutions can inform the extent to which the physics education research communities efforts address or perpetuate racism and sexism.

The results show a systemic gap in learning between White male physics students and their peers from marginalized groups. These gaps appear to be larger across races than across genders, yet the role of race/racism in college physics learning has received far less attention than gender/sexism (McCullough, 2018). The lack of attention to race/racism may have occurred because the low representation of students from marginalized racial groups in college physics courses made it difficult to establish statistically significant relationships between learning and race. Quantitative models can identify important differences in student outcomes but require major resources to identify causal relationships. As illustrated by the work on equity and gender in physics (Madsen, McKagan, & Sayre, 2013), even when models identify possible causes of inequities, they seldom provide a simple answer. Because of the power of quantitative data to drive policy, it is important that the models used to analyze the data are robust, accurate portrayals of reality. However, if a community sets standards so high that quantitatively investigating and redressing racism and sexism becomes infeasible, it is discarding objectivity and taking an unjust ideological stance. We propose taking an approach that acknowledges the limitations of datasets and statistical models while using them to identify and redress inequities.

**Limitations and future work**

In this investigation, we assumed that shifts in students' scores accurately measured their learning. This assumption ignores other factors that could influence students' scores such as their experiences and comfort taking tests, stereotype threat, language skills, how they interpret the questions, and biased questions. For example, McCaskey and colleagues (2004) found differences between what students said they personally believed and what they said scientists believed on the FCI and the disagreements were larger for women than for men. In another study, Traxler et al. (2018) identified gender bias on a handful of the items on the FCI. Henderson and colleagues (2018) did not find similar gender biases on the FMCE. Our pretest model found larger gender gaps on the FCI than the FMCE, which aligns with the findings from Traxler et al. (2018) and Henderson et al. (2018). Our gain models did not find differences in performance across the instruments. This is likely because of any biases having similar impacts on both the pretest and posttest, leaving the gain unaffected. While there are reasonable critiques of research-based assessments, the two assessments we used in this study have robust validation arguments created across several contexts and languages (e.g., Ishimoto et al., 2017).

In this investigation, we had the statistical power required to model the outcomes of students across genders and races/ethnicities. Our analysis showed the experience of students from various marginalized groups are not the same. As the number and types of LASSO users grow, we will further disaggregate marginalized populations within our models and examine how institutional contexts contribute to student inequities. For example, the learning outcomes of Black students at Historically Black Colleges and Universities, research intensive institutions, and 2-year colleges may vary in meaningful ways. By accounting for differences within groups across institutional contexts, researchers may find important trends in STEM student learning.

While we had the statistical power to model the differences across race and gender, we did not have the statistical power to identify differences in the relationship between collaborative instruction across races and genders. The small sample sizes for some groups in lecture, Pacific Islander and Black students, mean we cannot be certain that collaborative instruction equally benefits all groups of students. The education research community largely accepts that student-centered, collaborative instruction is always better for all students. Collaborative learning, however, is often used to mean many different things. For example, in our dataset we are not able to differentiate between an instructor who self-identifies their course as collaborative because they offer several opportunities for students to discuss a clicker question during their lecture from an instructor who self-identifies their course as collaborative because their students spend the majority of the course working in small groups on open ended problems. Resources need to be dedicated to identifying pedagogical practices and their impact on equity. Specifically, future research should examine whether the research-based strategies that have largely been developed at research-intensive, primarily White institutions align with the needs and resources for diverse students.

The LASSO database does not include information from students who choose not to complete the instruments and or choose not to share their data. Nissen et al., (2018) found that race and gender did not predict participation at the institution in their study. They found, however, that students with higher grades participated at higher rates. If the grade distributions in the courses in this study differ across races and genders, then the results in this study may be biased. We used hierarchical multiple imputation to minimize the amount of missing data and minimize the bias introduced by the missing data. However, the limitations of the dataset do not allow us to rule out the possibility that differences in participation rates across genders and races may have skewed the results. Only further studies can establish the reliability of our results.

## Acknowledgements

This work is funded in part by NSF-IUSE Grant No. DUE-1525338 and is Contribution No. LAA-052 of the Learning Assistant Alliance. The authors have no conflicts of interest.

# Supplemental material

## Descriptive statistics

Table S1
*Descriptive statistics for the sample, disaggregated by gender and race.*

| Race/Ethnicity | Instruction | Gender | N | $\mu_{pre}$ | $\sigma_{pre}$ | $\mu_{post}$ | $\sigma_{post}$ | $\mu_{gain}$ | $\sigma_{gain}$ |
|---|---|---|---|---|---|---|---|---|---|
| Asian | Collab. | Male | 811 | 36.1 | 22.3 | 55.3 | 26.6 | 19.2 | 20.6 |
| | | Female | 635 | 29.6 | 19.6 | 48.3 | 25.6 | 18.6 | 19.8 |
| | Lecture | Male | 96 | 33.8 | 20.5 | 53.9 | 24.2 | 20.1 | 21 |
| | | Female | 75 | 21.2 | 12.8 | 36.1 | 18.3 | 14.9 | 17.7 |
| Black | Collab. | Male | 169 | 27.9 | 15.5 | 44.1 | 21.6 | 16.2 | 19.1 |
| | | Female | 180 | 23 | 14.7 | 40.8 | 22.6 | 17.8 | 19.9 |
| | Lecture | Male | 28 | 22.7 | 12.2 | 43 | 18.9 | 20.3 | 16.9 |
| | | Female | 36 | 20.8 | 13.9 | 37.2 | 19.9 | 16.4 | 15 |
| Hispanic | Collab. | Male | 1192 | 30.7 | 18 | 49.8 | 24.2 | 19.1 | 19.3 |
| | | Female | 568 | 25.2 | 15.9 | 42.6 | 22.1 | 17.4 | 19 |
| | Lecture | Male | 204 | 26.6 | 12 | 40.5 | 19.5 | 13.9 | 18.7 |
| | | Female | 198 | 20.8 | 10.5 | 31.6 | 15.8 | 10.8 | 14.7 |
| Island | Collab. | Male | 69 | 35.2 | 17 | 55 | 21.9 | 19.8 | 17.9 |
| | | Female | 42 | 27.5 | 17.4 | 50.8 | 24.2 | 23.3 | 19.5 |
| | Lecture | Male | 5 | 23.2 | 10.2 | 41.3 | 35 | 18.1 | 28.1 |
| | | Female | 10 | 20.4 | 8.1 | 34 | 15.7 | 13.5 | 12 |
| Other | Collab. | Male | 497 | 33.7 | 21.3 | 51.8 | 24.9 | 18.1 | 19.2 |
| | | Female | 357 | 26.2 | 18.1 | 45.3 | 23.5 | 19.1 | 19.7 |
| | Lecture | Male | 92 | 27.8 | 17.2 | 42.2 | 22.3 | 14.4 | 21.4 |
| | | Female | 95 | 20.3 | 9.1 | 31.4 | 16.2 | 11.1 | 13.8 |
| White | Collab. | Male | 4791 | 44 | 21.7 | 66.1 | 24.1 | 22.1 | 19.2 |
| | | Female | 2429 | 31.3 | 17.3 | 54.2 | 23.4 | 22.9 | 19.5 |
| | Lecture | Male | 593 | 32.3 | 19.9 | 49 | 24.4 | 16.7 | 18.9 |
| | | Female | 685 | 21.1 | 11.9 | 37.9 | 20 | 16.8 | 18.2 |

## Assumption Checking

We are unaware of any simple method to pool the multi-level models created by multiply imputed datasets in a way that allows for a combined checking of assumptions. For this reason, we performed the assumption checks 10 times on our final model, once for each MI dataset. The assumption checks led to similar findings across each version of the final model. For the sake of simplicity, we only report the assumption check analysis of the final gain model from the first MI dataset. To test the assumption of linearity, we plotted the residual variance against the fitted values (Fig. S1a). In our visual inspection of Fig. S1a we saw no obvious trends and concluded that the model met the assumption of linearity. To test for homogeneity of variance we created a boxplot of the residuals across courses (Fig. S1b) and performed an ANOVA of the residuals across courses. A visual inspection of the boxplot showed the courses' residuals had consistent medians and interquartile ranges and therefore met the assumption of homogeneity of variance. The ANOVA supported our visual check because it did not find a statistically significant difference ($p>0.05$) in the variances across courses. Finally, we visually checked the assumption of normality of residuals using a quantile-quantile plot of the observed and expected values (Fig. S1c). We concluded that the model met the assumption of normality of residuals because of the linearity of the data in the plot.

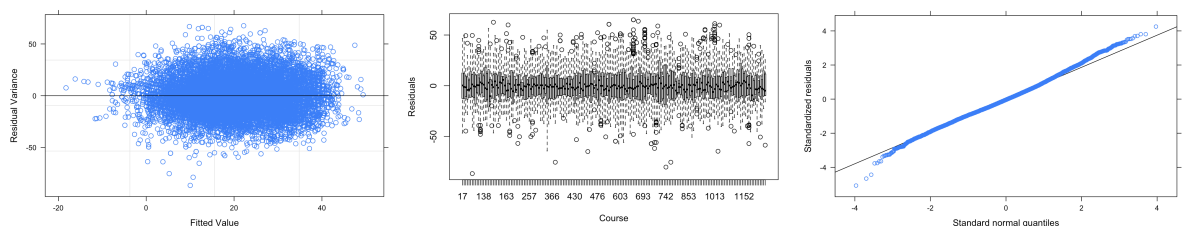

**Figure S1.** Figures to visually check three assumptions of HLM Model 6 from the first MI dataset. a) Assumption of linearity - A plot of residuals vs. fitted values. A random distribution of samples points indicated the model met the assumption of linearity. b) Assumption of homogeneity of variance - A boxplot of residuals by course. Consistent medians and interquartile ranges indicated the model met the assumption of homogeneity of variance. This assumption was further verified quantitatively using an ANOVA. c) Assumption of normality - A quantile-quantile plot of the observed and expected values. The points falling close to the line indicated that the model met the assumption of normality.

## Model Development

Table S2

*Pretest model development with final estimation of variance components (Var.) and the percent variance explained (% expl.) compared to Model 1, the unconditional model. Models 2 and 7 each failed to improve the percent explained variance by more than 1% at the two levels combined and the variables that were added in each were removed from the following model. Model 6 was identified as the simplest models that explained the most variance. Level 2 equations were combined into single rows when they were identical other than subscript numbers to simplify their presentation.*

| Model | Level | Equation | Var. | % expl. |
|---|---|---|---|---|
| 1 | 1 | $StudentPre_{ij} = \beta_{0j} + r_{ij}$ | 316.6 | - |
|   | 2 | $\beta_{0j} = \gamma_{00} + \mu_{0j}$ | 86.6 | - |
| 2 | 1 | $StudentPre_{ij} = \beta_{0j} + \beta_{1j} * \mathbf{Retake_{ij}} + r_{ij}$ | 315.8 | 0.30% |
|   | 2 | $\beta_{0j} = \gamma_{00} + \mu_{0j}$ <br> $\beta_{1j} = \gamma_{10}$ | 88.1 | -1.70% |
| 3 | 1 | $StudentPre_{ij} = \beta_{0j} + \beta_{1j} * + r_{ij}$ | 316.6 | 0% |
|   | 2 | $\beta_{0j} = \gamma_{00} + \gamma_{01} * \mathbf{FMCE_j} + \mu_{0j}$ <br> $\beta_{1j} = \gamma_{10}$ | 69.8 | 19.40% |
| 4 | 1 | $StudentPre_{ij} = \beta_{0j} + \beta_{1j} * \mathbf{Female_{ij}} + r_{ij}$ | 300.3 | 5.10% |
|   | 2 | $\beta_{0j} = \gamma_{00} + \gamma_{01} * FMCE_j + \mu_{0j}$ <br> $\beta_{1j} = \gamma_{10}$ | 58.7 | 23.20% |
| 5 | 1 | $StudentPre_{ij} = \beta_{0j} + \beta_{1j} * Female_{ij} +$ <br> $\beta_{2j} * \mathbf{Asian_{ij}} + \beta_{3j} * \mathbf{Black_{ij}} +$ <br> $\beta_{4j} * \mathbf{Hispanic_{ij}} + \beta_{5j} * \mathbf{PacIslander_{ij}} +$ <br> $\beta_{6j} * \mathbf{OtherRace_{ij}} + Female_{ij} * (\beta_{7j} * \mathbf{Asian_{ij}} +$ <br> $\beta_{8j} * \mathbf{Black_{ij}} + \beta_{9j} * \mathbf{Hispanic_{ij}} +$ <br> $\beta_{10j} * \mathbf{PacIslander_{ij}} + \beta_{11j} * \mathbf{OtherRace_{ij}}) + r_{ij}$ | 293.9 | 7.20% |
|   | 2 | $\beta_{0j} = \gamma_{00} + \gamma_{01} * FMCE_j + \mu_{0j}$ <br> $\beta_{(1-11)j} = \gamma_{(1-11)0}$ | 52.2 | 39.70% |
| 6 | 1 | $StudentPre_{ij} = \beta_{0j} + \beta_{1j} * Female_{ij} +$ <br> $\beta_{2j} * Asian_{ij} + \beta_{3j} * Black_{ij} +$ <br> $\beta_{4j} * Hispanic_{ij} + \beta_{5j} * PacIslander_{ij} +$ <br> $\beta_{6j} * OtherRace_{ij} + Female_{ij} * (\beta_{7j} * Asian_{ij} +$ <br> $\beta_{8j} * Black_{ij} + \beta_{9j} * Hispanic_{ij} +$ <br> $\beta_{10j} * PacIslander_{ij} + \beta_{11j} * OtherRace_{ij}) + r_{ij}$ | 293.7 | 7.20% |
|   | 2 | $\beta_{0j} = \gamma_{00} + \gamma_{01} * FMCE_j + \mu_{0j}$ <br> $\beta_{1j} = \gamma_{10} + \gamma_{11} * \mathbf{FMCE_j}$ <br> $\beta_{(2-11)j} = \gamma_{(2-11)0}$ | 51.4 | 40.70% |

| 7 | 1 | $StudentPre_{ij} = \beta_{0j} + \beta_{1j} * Female_{ij} + \beta_{2j} * Asian_{ij} + \beta_{3j} * Black_{ij} + \beta_{4j} * Hispanic_{ij} + \beta_{5j} * PacIslander_{ij} + \beta_{6j} * OtherRace_{ij} + Female_{ij} * (\beta_{7j} * Asian_{ij} + \beta_{8j} * Black_{ij} + \beta_{9j} * Hispanic_{ij} + \beta_{10j} * PacIslander_{ij} + \beta_{11j} * OtherRace_{ij}) + r_{ij}$ | 293.4 | 7.30% |
|---|---|---|---|---|
|   | 2 | $\beta_{0j} = \gamma_{00} + \gamma_{01} * FMCE_j + \mu_{0j}$<br>$\beta_{(1-6)j} = \gamma_{(1-6)0} + \gamma_{(1-6)1} * \mathbf{FMCE_j}$<br>$\beta_{(7-11)j} = \gamma_{(7-11)0}$ | 51.5 | 40.50% |

Table S3

*Gain model development with final estimation of variance components (Var.) and the percent variance explained (% expl.) compared to Model 1, the unconditional model. Models 4, 5, 9, and 10 each failed to improve the percent explained variance by more than 1% at the two levels combined and the variables that were added in each were removed from the following model. Models 7 and 8 were identified as the simplest models that explained the most variance. Model 7 predicts differences in student groups independent of the course type. Model 8 predicts differences in student groups while accounting of the course type. Level 2 equations were combined into single rows when they were identical other than subscript numbers to simplify their presentation.*

| Model | Level | Equation | Var. | % expl. |
|---|---|---|---|---|
| 1 | 1 | $Gain_{ij} = \beta_{0j} + r_{ij}$ | 325.2 | - |
| | 2 | $\beta_{0j} = \gamma_{00} + \mu_{0j}$ | 61.4 | - |
| 2 | 1 | $Gain_{ij} = \beta_{0j} + \beta_{1j} * \mathbf{StudentPreCen}_{ij} + r_{ij}$ | 295.3 | 9.10% |
| | 2 | $\beta_{0j} = \gamma_{00} + \mu_{0j}$ $\beta_{1j} = \gamma_{10}$ | 62.3 | -1.50% |
| 3 | 1 | $Gain_{ij} = \beta_{0j} + \beta_{1j} * StudentPreCen_{ij} + \beta_{2j} * \mathbf{Retake}_{ij} + r_{ij}$ | 294.7 | 9.40% |
| | 2 | $\beta_{0j} = \gamma_{00} + \mu_{0j}$ $\beta_{(1-2)j} = \gamma_{(1-2)0}$ | 59.8 | 2.60% |
| 4 | 1 | $Gain_{ij} = \beta_{0j} + \beta_{1j} * StudentPreCen_{ij} + \beta_{2j} * Retake_{ij} + r_{ij}$ | 294.7 | 9.40% |
| | 2 | $\beta_{0j} = \gamma_{00} + \gamma_{01} * \mathbf{CoursePreCen}_{j} + \mu_{0j}$ $\beta_{(1-2)j} = \gamma_{(1-2)0}$ | 60.2 | 2.00% |
| 5 | 1 | $Gain_{ij} = \beta_{0j} + \beta_{1j} * StudentPreCen_{ij} + \beta_{2j} * Retake_{ij} + r_{ij}$ | 294.7 | 9.40% |
| | 2 | $\beta_{0j} = \gamma_{00} + \gamma_{01} * \mathbf{FMCE}_{j} + \mu_{0j}$ $\beta_{(1-2)j} = \gamma_{(1-2)0}$ | 59.9 | 2.40% |
| 6 | 1 | $Gain_{ij} = \beta_{0j} + \beta_{1j} * StudentPreCen_{ij} + \beta_{2j} * Retake_{ij} + \beta_{3j} * \mathbf{Female}_{ij} + r_{ij}$ | 294.1 | 9.60% |
| | 2 | $\beta_{0j} = \gamma_{00} + \mu_{0j}$ $\beta_{(1-3)j} = \gamma_{(1-3)0}$ | 59.3 | 3.40% |
| 7 | 1 | $Gain_{ij} = \beta_{0j} + \beta_{1j} * StudentPreCen_{ij} + \beta_{2j} * Retake_{ij} + \beta_{3j} * Female_{ij} + \beta_{4j} * \mathbf{Asian}_{ij} + \beta_{5j} * \mathbf{Black}_{ij} + \beta_{6j} * \mathbf{Hispanic}_{ij} + \beta_{7j} * \mathbf{PacIslander}_{ij} + \beta_{8j} * \mathbf{OtherRace}_{ij} + Female_{ij} * (\beta_{9j} * Asian_{ij} + \beta_{10j} * Black_{ij} + \beta_{11j} * Hispanic_{ij} + \beta_{12j} * PacIslander_{ij} + \beta_{13j} * OtherRace_{ij}) + r_{ij}$ | 291.3 | 10.40% |
| | 2 | $\beta_{0j} = \gamma_{00} + \mu_{0j}$ $\beta_{(1-13)j} = \gamma_{(1-13)0}$ | 58.1 | 5.40% |
| 8 | 1 | $Gain_{ij} = \beta_{0j} + \beta_{1j} * StudentPreCen_{ij} + \beta_{2j} * Retake_{ij} + \beta_{3j} * Female_{ij} + \beta_{4j} * Asian_{ij} + \beta_{5j} * Black_{ij} + \beta_{6j} * Hispanic_{ij} + \beta_{7j} * PacIslander_{ij} + \beta_{8j} * OtherRace_{ij} + Female_{ij} * (\beta_{9j} * Asian_{ij} + \beta_{10j} * Black_{ij} + \beta_{11j} * Hispanic_{ij} + \beta_{12j} * PacIslander_{ij} + \beta_{13j} * OtherRace_{ij}) + r_{ij}$ | 291.3 | 10.40% |
| | 2 | $\beta_{0j} = \gamma_{00} + \gamma_{01} * \mathbf{Lecture}_{j} + \mu_{0j}$ | 57 | 7.20% |

| 9 | 1 | $\beta_{(1-13)j} = \gamma_{(1-13)0}$<br><br>$Gain_{ij} = \beta_{0j} + \beta_{1j} * StudentPreCen_{ij} + \beta_{2j} * Retake_{ij} + \beta_{3j} * Female_{ij} + \beta_{4j} * Asian_{ij} + \beta_{5j} * Black_{ij} + \beta_{6j} * Hispanic_{ij} + \beta_{7j} * PacIslander_{ij} + \beta_{8j} * OtherRace_{ij} + Female_{ij} * (\beta_{9j} * Asian_{ij} + \beta_{10j} * Black_{ij} + \beta_{11j} * Hispanic_{ij} + \beta_{12j} * PacIslander_{ij} + \beta_{13j} * OtherRace_{ij}) + r_{ij}$ | 291.4 | 10.40% |
|---|---|---|---|---|
|   | 2 | $\beta_{0j} = \gamma_{00} + \gamma_{01} * Lecture_j + \mu_{0j}$<br>$\beta_{(1-2)j} = \gamma_{(1-2)0}$<br>$\beta_{(3-13)j} = \gamma_{(3-13)0} + \gamma_{(3-13)1} * \textbf{\textit{Lecture}}_j$ | 57.1 | 7.00% |